\documentclass[preprint,showpacs,preprintnumbers,amsmath,amssymb, superscriptaddress,longbibliography,nofootinbib,showkeys]{revtex4-1}

\usepackage{textcomp}

\usepackage{makeidx}
\usepackage{amsmath}
\usepackage{subfig}
\usepackage{amssymb}
\usepackage{hyperref}
\usepackage{graphicx}
\usepackage{epstopdf}
\usepackage{color}
\usepackage{soul}

\begin{document}

\title{Representation of compact stars using the black string set--up}

\author{Milko Estrada} \email[]{milko.estrada@ua.cl}

\affiliation{Facultad de Ingenier\'ia, Ciencia y Tecnolog\'ia, Universidad Bernardo O'Higgins, Santiago, Chile }

\author{Francisco  Tello-Ortiz} \email[]{francisco.tello@ua.cl}
\affiliation{Departamento de F\'isica, Facultad de Ciencias B\'asicas, Universidad de Antofagasta, Casilla 170, Antofagasta, Chile.}

\author{Ksh. Newton Singh}
\email[]{ntnphy@gmail.com}
\affiliation{Department of Physics, National Defence Academy, Khadakwasla, Pune-411023, India.}

\author{S. K. Maurya}
\email[]{sunil@unizwa.edu.om}
\affiliation{Department of Mathematical and Physical Sciences,
College of Arts and Sciences, University of Nizwa, Nizwa, Sultanate of Oman.}

\date{\today}

\begin{abstract}
This work is devoted to provide a way of representing $4D$ spherically symmetric and static compact stellar configurations into a $5D$ space time, using the black string framework. 

We write the four and five dimensional line elements and the four and five dimensional energy momentum tensor such that the four and five dimensional quantities are related by a function $A(z)$, where $z$ represents to the extra dimension. Remarkably, one consequence of our chosen form for the line element, for the energy momentum tensor and for $A(z)$, is the fact that the $5D$ equations are reduced to the usual form of the four dimensional equations of motion. Also, the five dimensional conservation equation adopts the form of the four dimensional conservation equation. It is worth mentioning that, although our methodology is simple, this form of reduction could serve to represent other types of four dimensional objects into an extra dimension in future works. Furthermore, under our assumptions the sign of the pressure along the extra dimension is given by the trace of the four dimensional energy momentum tensor.

Furthermore, our simple election for the function $A(z)$ modifies some features of the induced 4--dimensional compact stellar configuration, such as the mass--radius relation, the momentum of inertia, the central values of the thermodynamic variables, to name a few.  
This clearly motivates the study of the impact of this methodology on some particular 4--dimensional models. In this concern, we have considered as a toy model the well--known isotropic Buchdahl model, showing by a graphical analysis how the mentioned properties are altered. It is worth mentioning that the resulting solution is well--behaved, satisfying all the physical criteria to represent a toy physical stellar model. Besides, the topology of this model is ${S}^{2}\times {S}^{1}$ and not the ${S}^{3}$ topology of the 5--dimensional structures.
\end{abstract}

\maketitle

\section{Introduction}

Along last years, several branches of theoretical physics have predicted the existence of extra dimensions. 
Supporting this idea, we have the well--known string theory, higher dimensional black holes and higher dimensional stellar distributions, to name a few.

In this concern, higher dimensional stellar distributions have been widely studied in the literature. See for example references \cite{Chilambwe:2015rra, Dadhich:2016wtb,Molina:2016xeu,Bhar:2014tqa,Paul:2018vho,Estrada:2018vrl}. It is worth to mention that in all these references the topology of a $D$--dimensional stellar distributions corresponds to $S^{D-2}$ topology. For example, the topology of a five dimensional stellar distribution corresponds to a $S^3$ topology. However, it is still an open problem to understand how to represent $4D$ spherically symmetric and static compact stellar distributions within the framework of extra dimensions, where the topology should differ from the $S^3$ topology.

\begin{figure}
\centering
\includegraphics[scale=0.85]{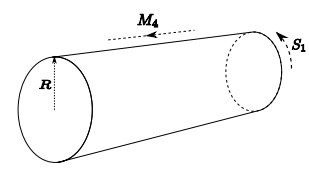}
\caption{KK set-up \cite{Shifman:2009df}\label{FigKK}}.
\end{figure}

\begin{figure}
\centering
\includegraphics[scale=0.85]{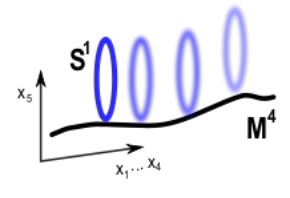}
\caption{KK set-up \label{FigKK1}}.
\end{figure}

One the the first models in representing $4D$ geometries, immersed in extra dimensions was the well--known  Kaluza--Klein (KK) model. In this model, the higher dimensional space--time representation corresponds to
the product between the Minkowski space--time and a compactified extra dimension. The most common example is $\mathcal{M}_{4}\times S^{1}$, being $\mathcal{M}_{4}$ the Minkowski manifold and $S^{1}$ a sphere of radius $\mathcal{R}$. In this case, $\mathcal{M}_{4}$ corresponds to the horizontal direction, whilst the sphere $S^{1}$ is the boundary. Therefore, the whole space--time is a cylinder (see figure \ref{FigKK} of reference \cite{Shifman:2009df} and figure \ref{FigKK1}).

\begin{figure} 
\centering
\includegraphics[scale=0.85]{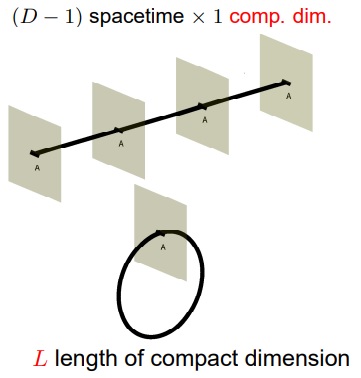}
\caption{BS set-up \cite{Kleihaus}\label{FigSlides}}.
\end{figure}

Other models admitting a $(D-1)$--dimensional geometry representation immersed in extra dimensions, is the so--called black strings (BS). This scenario, is the simplest extension of black hole solutions. Specifically, BS consists in slice replications of a $(D-1)$--dimensional black hole along a compact extra dimension (see for example figure \ref{FigSlides}). It is worth mentioning that the BS setup, is different from the KK mechanism, because the KK approach includes more degrees of freedom (vector and scalar fields) in addition to the gravitational ones.

\begin{figure}
\centering
\includegraphics[scale=0.65]{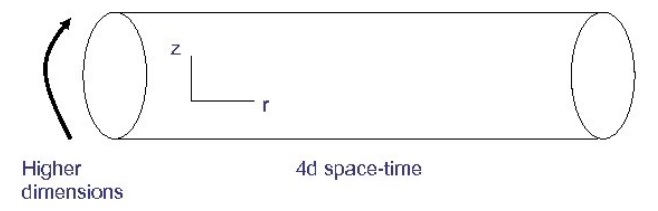}
\caption{Cylindrical set--up of BS \label{FigCilindro}}.
\end{figure}

In the simplest BS scenario (the case $D=5$), the radial coordinate of a 4--dimensional black hole corresponds to the horizontal direction, while the extra coordinate corresponds to the boundary of a $S^{1}$ sphere. The reasons why this is the simplest scheme, lie on the fact that the horizontal direction always coincides with the 4--dimensional space--time and not only with the radial component. So, this scenario corresponds to the situation shown by figure \ref{FigCilindro}.  

\begin{figure}
\centering
\includegraphics[scale=0.85]{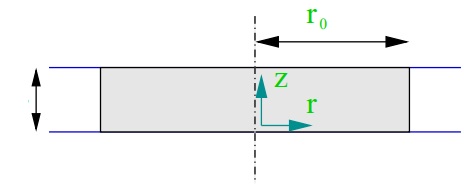}
\caption{KK BS \cite{Kol:2004ww} \label{FigKKBS}}.
\end{figure}

The simplest version of BS is given by the line element 
\begin{equation} \label{Kalusa}
    ds_{5D}^2=d{s}_{4D}^2+ dz^2,
\end{equation}
 where $d{s}_{4D}^2$ is the 4--dimensional Schwarzschild space--time. The line element \eqref{Kalusa} corresponds to the Kaluza--Klein black string (KKBS). This solution has $S^2 \times R$ topology for non compact $z$. The representation of this BS for a compact extra coordinate $z$, can be visualized in the figure \ref{FigKKBS} (see reference \cite{Kol:2004ww} for further details). In this case, the topology corresponds to the product between the horizontal direction, whose topology is $S^2$ and the compact extra dimension, with topology $S^1$, {\it i.e.} the topology is $S^2 \times S^1$. So, for $r<r_0$, being $r_0$ the Schwarzschild radius, the interior region is denominated as {\it cylindrical event horizon}. So, as was claimed in reference \cite{Kleihaus:2016kxj}, the horizon completely wraps the compact extra dimension. 
 
Following figures \ref{FigKK}, \ref{FigKK1} and \ref{FigSlides}, similar slices of the 4--dimensional surface $(t,r,\theta,\phi)$ (see figures \ref{FigCilindro} and \ref{FigKKBS}, for a simplified point of view with similar radial points $r$) are replicated in different locations of the compact extra dimension $z$. In this respect, in the reference \cite{Kleihaus:2006ee} was argued that the KKBS solution shows translational symmetry along the extra coordinate direction. 
 
 It is worth mentioning that in the KKBS the 4--dimensional Schwarzschild black hole is represented (whose topology is $S^2$) into a compact extra dimension (whose topology is $S^1$). So, the $S^2 \times S^1$ topology of the KKBS differs from the usual $S^3$ topology of the 5--dimensional Schwarzschild--Tangherlini solution \cite{Kol:2002xz}. 
 On the other hand, in the KKBS the topology of the central singularity corresponds to the product between the Schwarzschild singularity and the $S^1$ manifold for a compact extra coordinate.
 
 Recently, a new methodology has been provided for the representation of a 4--dimensional regular black hole into an extra dimension using the black string framework \cite{Estrada:2021naf}. An interesting extension of this work was developed in \cite{Culetu:2021zun}. So, motivated by this methodology, here the main aim of the present work, is to test the possibility to carry out the representation of a 4--dimensional compact stellar distribution into an extra dimension, using the black strings setup. So, instead the usual $S^3$ topology of the 5--dimensional stellar distributions, the topology of our solution, which corresponds to a 4--dimensional spherically symmetric and static stellar configuration into an extra dimension, is $S^2 \times R(S^1)$ for a non compact extra dimension. 

So, in this article we provide a specific form for the 4--dimensional and 5--dimensional line elements, along with a suitable form for the energy--momentum tensor. We start from the 5--dimensional Einstein's field equations and, after a suitable choice on the function $A(z)$, which deforms the 4--dimensional geometry, the 5--dimensional field equations are remarkably reduced to the usual form of the 4--dimensional equations of motion.
Moreover, as toy a model, we apply the mentioned framework to the well--known  Buchdahl type geometry \cite{Buchdahl:1959zz} as 4--dimensional manifold, representing the compact stellar distribution.

\section{The Model} \label{sec3}

In this section we present in brief the strategy to represent a 4--dimensional manifold in a 5--dimensional space--time, by employing the BS framework. The underlying 4--dimensional geometry will represent to a stellar distribution. The starting point are the 5--dimensional Einstein's field equations given by 
\begin{equation} \label{EcuacionDeEinstein}
    G^M_{N}+\Lambda_{5D} \delta^M_{N}= 8 \pi \tilde{G}_{5D} T^M_{N},
\end{equation}
where $M,N=0,1,2,3,5$ and $\tilde{G}_{5D}$ is the 5--dimensional Newton's constant. The line element is:
\begin{align}
ds_{5D}^2=& W(z) \cdot d\bar{s}_{4D}^2+ dz^2, \label{elementodelinea} \\
=& e^{-2A(z)} \cdot g_{\mu \nu}^{(4D)} dx^\mu dx^\nu + dz^2, \label{elementodelineaIntermedio} \\
=& e^{-2A(z)}  \left[-e^{\nu(r)}dt^2+\mu^{-1}(r)dr^2+r^2 d\Omega^2 \right]+ dz^2, \label{Elementodelinea1}
\end{align}
with $\mu,\nu=0,1,2,3$. In order to have a regular 5--dimensional geometry \cite{Estrada:2021naf}, the extra coordinate $z$ should be compact at the BS style, in sense as was explained before. In this work we consider $z \in [-\pi,\pi]$. Besides, the function $W$ depends only on the extra coordinate $z$. Thus, the 5--dimensional geometry is modified by the function $W(z)$. Although, similar slices of the 4--dimensional surface are replicated along the extra dimension $z$, the 5--dimensional geometry no longer has traslational symmetry along the $z$ coordinate, as occurs in the KKBS \cite{Kleihaus:2006ee}.

Along with previous considerations, here we propose the following energy--momentum tensor
\begin{equation} \label{EnergiaMomentum1}
    T^M_{N}=\mbox{diag}\big (-\rho(r,z),p_r(r,z),p_\theta(r,z),p_\phi(r,z),p_z(r,z) \big),
    \end{equation}
where $\rho$ denotes the energy density while, $p_r$, $p_\theta$, $p_\phi$ and $p_z$ are representing the pressure components. Some examples in $5D$, where the energy--momentum tensor depends on the extra coordinate can be found at \cite{Culetu:2021zun}, and energy--momentum tensors depending on the radial coordinate in a BSs scenario surrounded by quintessence matter can be found at \cite{Ali:2019mxs}.    

\subsection{Relation between $4D$ and $5D$ Newton's constants}

In order to establish a relation between the 4--dimensional and $D$--dimensional Newton's constants, we are going to follow the approach given in: ``heuristics of higher-dimensional gravity'' \cite{Maartens:2010ar}. Then we will see that this analysis serves to reduce the 5--dimensional field equations to 4--dimensions. In this reference, for a $D$ dimensional space--time, the effective four dimensional Planck scale $M_4$ is given by
\begin{equation} \label{Constantes1}
    M_4^2=M_D^{D-2}L^{D-4},
\end{equation}
where $M_D$ represents the $D$ dimensional Planck scale and where $L$ has units of length. In this context, as indicates the reference \cite{Maartens:2010ar}, the Newton constant in $D$ dimensions is given by:

\begin{equation} \label{Constantes2}
    \tilde{G}_D = \frac{1}{M_D^{D-2}}.
\end{equation}

So, equation \eqref{Constantes1} can be re--written as:
\begin{equation} \label{Constantes3}
    \tilde{G}_D=\tilde{G}_4 L^{D-4}.
\end{equation}

Using Planck units, the $D$ dimensional Newton's constant has the following units $[\tilde{G}_D]=\ell_p^{D-2}$ \cite{Garraffo:2008hu}, where $\ell_p$ corresponds to the Planck length. Also, this can be obtained from equation \eqref{Constantes2}, assuming that $[M_D]=\ell_p^{-1}$. Besides, using the same system of units in equation \eqref{Constantes1}, one gets $[L]=\ell_p$.

From equation \eqref{Constantes3}:
\begin{equation} \label{Constantes31}
    \tilde{G}_5=\tilde{G}_4 L.
\end{equation}

Since the units of the Ricci scalar and the Einstein tensor are $[R]=[G^M_N]=\ell_p^{-2}$ and $[\sqrt{g}d^Dx]=\ell_p^D$, respectively. Then, the gravitational action is dimensionless $[\tilde{G}_D^{-1} \sqrt{g}d^dxR]=\ell_p^0$.  

Following \cite{Aros:2019quj}, the energy--momentum tensor has units of $[T^M_N]=\ell_p^{-D}$, and then, $[G^M_N]=[\tilde{G}_D \cdot T^M_N]=\ell_p^{-2}$.

So, in order to describe the 5--dimensional energy--momentum tensor in terms of 4--dimensional quantities, we define a 4--dimensional energy--momentum as $\bar{T}^\mu_\nu$, where $[\bar{T}^\mu_\nu]=\ell_p^{-4}$, and constant $K$ with dimensions $[K]=\ell_p^{4-D}$, such that:

\begin{equation} \label{Constantes4}
    [\bar{T}^\mu_{\nu}]=[\delta^\mu_M \delta^N_\nu \cdot K^{-1} \cdot T^M_{N}]=\ell_p^{-4}
\end{equation}
{\it i.e.}, for $D=5$, $[K]=\ell_p^{-1}$ and $[T^M_{N}]=\ell_p^{-5}$.

\subsection{Equations of motion}

For the line element \eqref{Elementodelinea1} and the energy--momentum tensor \eqref{EnergiaMomentum1}, Einstein's field equations are:

\begin{eqnarray}\label{EqMovimiento5D}
&&\hspace{-0.5cm} 6(\dot{A})^2-3 \ddot{A} + \Lambda_{5D}+ e^{2A(z)}\bigg\{\frac{\mu'}{r}-\frac{1}{r^2}+\frac{\mu}{r^2}\bigg\}=- 8 \pi \tilde{G}_{5D} \rho(r,z), \label{1tt} \\ 
&&\hspace{-0.5cm} 6(\dot{A})^2-3 \ddot{A}+ \Lambda_{5D}+e^{2A(z)}\bigg\{\frac{\mu \nu'}{r}-\frac{1}{r^2}+\frac{\mu}{r^2}\bigg\}=8 \pi \tilde{G}_{5D} p_r (r,z), \label{1rr} \\ 
&&\hspace{-0.5cm} 6(\dot{A})^2-3 \ddot{A} + \Lambda_{5D}+e^{2A(z)}\bigg\{ \frac{\mu \nu'}{2r}+\frac{\mu'}{2r} +\frac{1}{4}\mu'\nu'+\frac{1}{2}\mu \nu''+\frac{1}{4}\mu (\nu')^2 \bigg\}=8 \pi \tilde{G}_{5D} p_\theta (r,z), \label{1thetatheta} \\
&&\hspace{-0.5cm} 6(\dot{A})^2+\Lambda_{5D}+e^{2A(z)}\bigg\{  \frac{1}{4}\mu'\nu'+\frac{1}{2}\mu \nu''+\frac{1}{4}\mu (\nu')^2+\frac{\mu \nu'}{r}+\frac{\mu'}{r}-\frac{1}{r^2}+\frac{\mu}{r^2} \bigg\}=8 \pi \tilde{G}_{5D} p_z (r,z),~~~~~~\label{1zz}
\end{eqnarray}
where dot indicates derivation with respect to the extra coordinate $z$ and the primes represent derivation with respect to the radial coordinate $r$.

Following the same ansatz for $A(z)$ given in \cite{Estrada:2018vrl}, one obtains

\begin{equation} \label{1EcuacionA}
        6(\dot{A})^2 =- \Lambda_{5D},
    \end{equation}
with 
\begin{equation}
    \Lambda_{5D}=-\frac{6}{l^2},
\end{equation}
then
\begin{equation} \label{warpfactor1}
e^{-2A(z)}=C e~^{\mp z/l},
\end{equation}
where $C$ a positive and dimensionless constant and, $l$ corresponds to the AdS radius when the space--time corresponds to an AdS space--time. For the sake of simplicity, we consider that the magnitude of $C=1$. Therefore, for this particular case we will consider the positive sign in equation \eqref{warpfactor1}.

Using the ansatz \eqref{1EcuacionA}, the  three first terms in left member of Eqs. \eqref{1tt}, \eqref{1rr} and \eqref{1thetatheta}, and the two first terms in the left  member of the equation \eqref{1zz}  are cancel--out.

In order to write the equations of motion in terms of the $4D$ Newton's constant, we use the relation \eqref{Constantes3}. Thus, Einstein's field equations are:

\begin{eqnarray}\label{EqMovimiento5D}
&&\hspace{-0.5cm}  W^{-1}\bigg\{\frac{\mu'}{r}-\frac{1}{r^2}+\frac{\mu}{r^2}\bigg\}=- 8 \pi L \tilde{G}_4 \rho(r,z), \label{2tt} \\ 
&&\hspace{-0.5cm} W^{-1}\bigg\{\frac{\mu \nu'}{r}-\frac{1}{r^2}+\frac{\mu}{r^2}\bigg\}=8 \pi L \tilde{G}_4 p_r (r,z), \label{2rr} \\ 
&&\hspace{-0.5cm}W^{-1}\bigg\{ \frac{\mu \nu'}{2r}+\frac{\mu'}{2r} +\frac{1}{4}\mu'\nu'+\frac{1}{2}\mu \nu''+\frac{1}{4}\mu (\nu')^2 \bigg\}=8 \pi L \tilde{G}_4 p_\theta (r,z), \label{2thetatheta} \\
&&\hspace{-0.5cm} W^{-1}\bigg\{  \frac{1}{4}\mu'\nu'+\frac{1}{2}\mu \nu''+\frac{1}{4}\mu (\nu')^2+\frac{\mu \nu'}{r}+\frac{\mu'}{r}-\frac{1}{r^2}+\frac{\mu}{r^2} \bigg\}=8 \pi L \tilde{G}_4 p_z (r,z),~~~~~~\label{2zz}
\end{eqnarray}

Besides, we propose the following energy--momentum tensor (\ref{EnergiaMomentum1}) 
\begin{align} \label{EnergiaMomentum2}
T^M_{N} =& F \big ( W  (z) \big) \cdot K \cdot \mbox{diag}[-\bar{\rho}(r),\bar{p}_r(r),\bar{p}_\theta(r),\bar{p}_\phi(r),\hat{p}_z(r)],
\end{align}
where $F \big ( W  (z) \big)$ represents a function depending on the extra coordinate and $[K]=\ell_p^{-1}$ in order to be consistent with the equation \eqref{Constantes4}.

For the sake of simplicity, the above energy--momentum tensor can be written as follows
\begin{equation} \label{EnergiaMomentum}
    T^M_{N}=\delta^M_\mu \delta^\nu_N \cdot F \big ( W  (z) \big) \cdot K \cdot \bar{T}^\mu_{\nu} + \delta^{M}_{5} \delta^{5}_{N} T^5_{5},
\end{equation}
As we will see in the equation \eqref{EcuacionReducida}, $\bar{T}^\mu_{\nu}$ represents the energy--momentum tensor associated to the 4--dimensional geometry, given by
\begin{equation} \label{EnergiaMomentum4D}
    \bar{T}^\mu_{\nu}=   \mbox{diag}(-\bar{\rho}(r),\bar{p}_r(r),\bar{p}_\theta(r),\bar{p}_\phi(r))
\end{equation}
Thus $\bar{\rho}(r),\bar{p}_r(r),\bar{p}_\theta(r),\bar{p}_\phi(r)$ are representing the energy density and the $r$, $\theta$, $\phi$ pressure components of the 4--dimensional geometry.

It is worth to mentioning that in our case, the 4--dimensional geometry is stretched along the extra dimension, then the 4--dimensional matter given by  $\bar{T}^\mu_\nu$ spread over the whole extra dimension. So, the value of the 5--dimensional energy--momentum tensor varies for different values of $z$ through the function $F \big ( W  (z) \big)$. This is also consistent with the fact that although similar slices of the 4--dimensional surface are replicated along the extra dimension $z$, the 5--dimensional geometry no longer has traslational symmetry along the $z$ coordinate, as occurs for the KKBS \cite{Kleihaus:2006ee}.

On the other hand, $T^5_{5}$ corresponds to the energy--momentum tensor associated to the extra dimension
\begin{equation} \label{EnergiaMomentum5}
    T^5_{5}= \delta^5_5  F \big ( W  (z) \big) K \hat{p}_z(r),
\end{equation}
where $\hat{p}_z(r)$ represents the radial dependency of $T^5_{5}$.

The components $\rho$, $p_r$, $p_\theta$, $p_\phi$, $p_z$ of the 5--dimensional energy--momentum tensor, are given in equation \eqref{EnergiaMomentum2}, such that the components of the 4--dimensional energy--momentum tensor $\bar{\rho}$, $\bar{p}_r$, $\bar{p}_\theta$, $\bar{p}_\phi$ have units of $[\bar{T}^\mu_\nu]=\ell_p^{-4}$ (for further details see equation \eqref{Constantes4}).

At this point it is important to stress that the function $W(z)$ modifies both, the geometry and the components of the energy--momentum tensor. 

It is worth mentioning that, in an arbitrary way and for sake of simplicity, we impose that constants $L$ and $K$ have magnitude equal to the unity, $L=K=1$. Moreover, hereinafter we shall consider $\tilde{G}_5=\tilde{G}_4=1$ \cite{Alcubierre:2018ahf,Spallucci:2021ljr}.

Now, for the line element \eqref{Elementodelinea1} and the energy--momentum tensor \eqref{EnergiaMomentum} the units of $[L \cdot K]= \ell_p^0$ are cancelled. Thus, the $(t,t)$, $(r,r)$, $(\theta,\theta)$ and $(z,z)$ components of the field equations are given by
\begin{eqnarray}\label{EqMovimiento5D}
&&\hspace{-0.5cm} W^{-1}\bigg\{\frac{\mu'}{r}-\frac{1}{r^2}+\frac{\mu}{r^2}\bigg\}=- 8 \pi F(W) \bar{\rho}, \label{3tt} \\ 
&&\hspace{-0.5cm} W^{-1}\bigg\{\frac{\mu \nu'}{r}-\frac{1}{r^2}+\frac{\mu}{r^2}\bigg\}=8 \pi F(W) \bar{p}_r, \label{3rr} \\ 
&&\hspace{-0.5cm} W^{-1}\bigg\{ \frac{\mu \nu'}{2r}+\frac{\mu'}{2r} +\frac{1}{4}\mu'\nu'+\frac{1}{2}\mu \nu''+\frac{1}{4}\mu (\nu')^2 \bigg\}=8 \pi F(W) \bar{p}_\theta, \label{3thetatheta} \\
&&\hspace{-0.5cm} W^{-1}\bigg\{  \frac{1}{4}\mu'\nu'+\frac{1}{2}\mu \nu''+\frac{1}{4}\mu (\nu')^2+\frac{\mu \nu'}{r}+\frac{\mu'}{r}-\frac{1}{r^2}+\frac{\mu}{r^2} \bigg\}=8 \pi F(W) \hat{p}_5,~~~~~~\label{3zz}
\end{eqnarray}
where we impose that:
\begin{equation} \label{EcuacionW}
    F \left(W (z) \right) = W^{-1}= e^{2A(z)}.
\end{equation}

Thus, by means of equations \eqref{1EcuacionA} and \eqref{EcuacionW} it is straightforward to check that the components \eqref{3tt}, \eqref{3rr} and \eqref{3thetatheta} of the 5--dimensional equations of motion are reduced to the usual ones corresponding to the 4--dimensional geometry $d\bar{s}_{4D}^2$. This latter is given by equations \eqref{elementodelinea}, \eqref{elementodelineaIntermedio} and \eqref{Elementodelinea1}
\begin{eqnarray}\label{EqMovimiento5D1}
&&\frac{\mu'}{r}-\frac{1}{r^2}+\frac{\mu}{r^2}=- 8 \pi  \bar{\rho} \label{tt1} \\ 
&&\frac{\mu \nu'}{r}-\frac{1}{r^2}+\frac{\mu}{r^2}=8 \pi  \bar{p}_r, \label{rr1} \\ 
&& \frac{\mu \nu'}{2r}+\frac{\mu'}{2r} +\frac{1}{4}\mu'\nu'+\frac{1}{2}\mu \nu''+\frac{1}{4}\mu (\nu')^2 =8 \pi \bar{p}_\theta. \label{thetatheta1} 
\end{eqnarray}

Thus, the above equations have the following form 
\begin{equation} \label{EcuacionReducida}
\bar{G}^\mu_\nu \propto \bar{T}^\mu_\nu.
\end{equation}

An important consequence in employing the BS setup, lies on the fact that using the line element \eqref{Elementodelinea1}, for the functions \eqref{warpfactor1} \eqref{EcuacionW}, the $(\mu,\nu)$ components of the 5--dimensional equations of motion adopt the usual 4--dimensional ones \eqref{EcuacionReducida}, corresponding to the line element $d\bar{s}_{4D}^2$ .

On the other hand, a direct computation reveals that the $z$--$z$ component, equation \eqref{3zz}, takes the form
\begin{equation}
     \bar{p}_\theta- \frac{\bar{\rho}}{2}+\frac{\bar{p}_r}{2}=\hat{p}_5,
\end{equation}

In order to test the sign of the pressure along the extra dimension, $\hat{p}_5$, we rewrite the last equation as:
\begin{equation} \label{ecuacionp5}
2 \hat{p}_5 = - (\bar{\rho}-\bar{p}_r-2\bar{p}_\theta)=\bar{T}.
\end{equation}

Thus, under our assumptions, the sign of the pressure $\hat{p}_5$ is given by the sign of the trace of the 4--dimensional energy--momentum tensor $\bar{T}^{\mu\nu}$. This is known as the Trace Energy Condition (TEC), and should be negative when the signature is $(-,+,+,+)$ \cite{Martin-Moruno:2017exc}. However, as mentioned in \cite{Martin-Moruno:2017exc}, the study of TEC has been abandoned after the 70s decade.

It is worth to mentioning that, for the chosen energy--momentum \eqref{EnergiaMomentum2}, along with the line element \eqref{Elementodelinea1}, the relation \eqref{EcuacionW} and, for the present case where $\bar{p}_\theta=\bar{p}_\phi$, the conservation equation $\nabla_M T^M_N=0$ is:
\begin{eqnarray}\label{tov}
&& \hspace{-1cm} - \frac{1}{2} \dot{F} \bigg [  2 \hat{p}_5 +  \Big(\bar{\rho} - \bar{p}_r- 2\bar{p}_\theta \Big)   \bigg ] \nonumber \\
&& - F \bigg[ - \frac{\nu'}{2} (\bar{\rho}+\bar{p}_r) - \frac{d \bar{p}_r}{dr} +\frac{2}{r} (\bar{p}_\theta-\bar{p}_r) \bigg] =0.
\end{eqnarray}

The condition given by equation \eqref{ecuacionp5}, allows us to eliminate the first bracket in the left hand side of (\ref{tov}). On the other hand, since the extra coordinate $z$ is a compact coordinate, from the condition \eqref{EcuacionW} the factor $F \neq 0$, hence the conservation equation adopts the form of the usual 4--dimensional conservation law 
\begin{equation}
- \frac{\nu'}{2} (\bar{\rho}+\bar{p}_r) - \frac{d \bar{p}_r}{dr} +\frac{2}{r} (\bar{p}_\theta-\bar{p}_r) =0 .
\end{equation}

\begin{figure}
\centering
\includegraphics[scale=0.65]{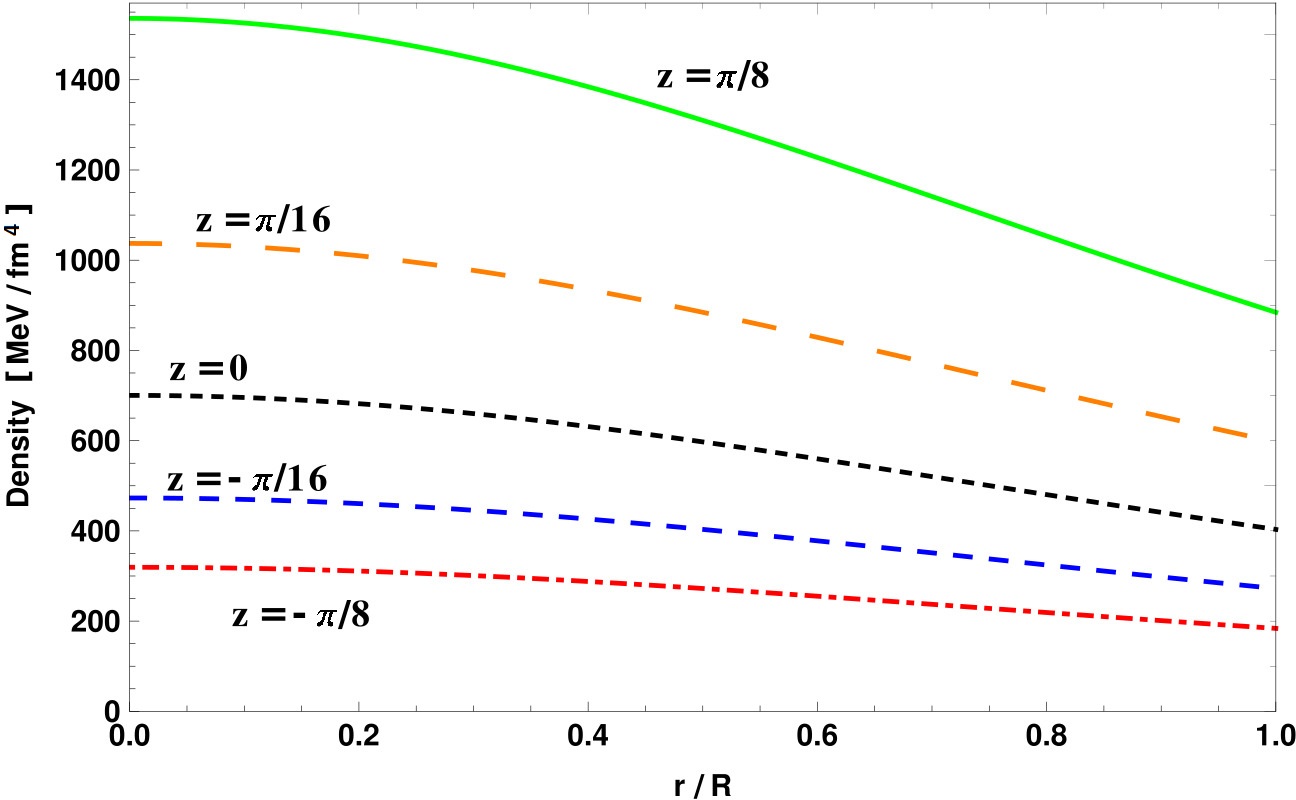}\\
\includegraphics[scale=0.65]{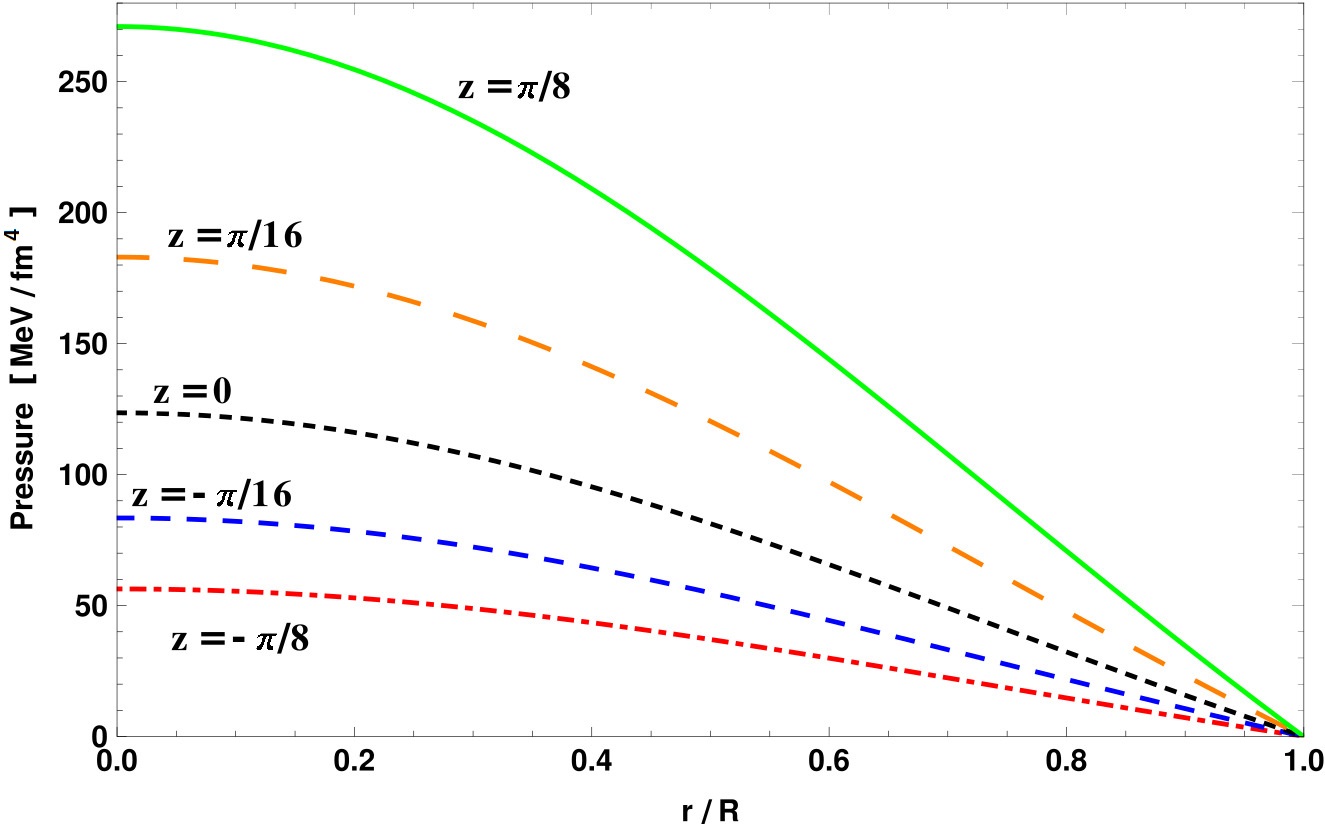}
\caption{\textbf{Upper panel}: The trend of the density against the dimensionless radial coordinate $r/\bar{R}$. \textbf{Lower panel}: The behavior of the pressure versus $r/\bar{R}$. These plots consider that the extra coordinate $z$ is taking values in the interval $[-\pi,\pi]$. }\label{f1}
\end{figure}

It is remarkable to note that, although the present form to solve the equations of motion is simple in its right own, this way allows that the 5--dimensional equations of motion are  reduced to the usual Einstein's 4--dimensional field equations. Therefore, this methodology could serve to represent other types of 4--dimensional objects into an extra dimension in future works.

\subsection{About the five dimensional matching conditions} \label{Matching}

We assume that the 4--dimensional geometry is stretched between $z \in [-\pi,\pi]$. So, we call $z_0$ to each point belonging to the domain of the extra coordinate $z$. 

In to order to find all constant parameters characterizing the model, we shall employ the well--known Israel--Darmois junction conditions. Those are given by

\begin{itemize}
   \item The 4--dimensional line element must be continuous,{\it i.e} at the exterior of the star the geometry must be described by the exterior Schwarzschild space--time. This is consistent with the fact that the induced metric, for all point $z_0 \in z$, where is stretched the $4D$ geometry along the extra dimension, is given by \cite{Aros:2012xi}
 \begin{equation}   \label{metricIanducida}
    \tilde{g}_{\mu \nu} = g_{\mu \nu}^{(4D)} W(z_0).
\end{equation}

So, at the boundary of the star, $r=\bar{R}$, for a fixed value $z=z_0$, the internal geometry of the star \eqref{elementodelinea} is
\begin{equation}
     (ds_{5D}^2)^-= W(z_0) \cdot d\bar{s}_{4D}^2 ~~~(\mbox{Star}).
     \end{equation}
It must be matched with
\begin{equation}
(ds_{5D}^2)^+= W(z_0) \cdot d{s}_{4D}^2 ~~~(\mbox{Schwarzschild}).
\end{equation}
So, the induced line element must be continuous {\it i.e.} there must be no jumps in the metric:
\begin{equation} \label{MatchingStar}
    (g_{\mu \nu}^{(4D)})^-_{\mbox{star}}=(g_{\mu \nu}^{(4D)})^-_{\mbox{Schw}}.
\end{equation}
Thus, inside the stellar distribution, the topology corresponds to the product between the 4--dimensional star and $S^1$. However, outside the stellar distribution, the topology corresponds to the product between the Schwarzschild geometry and the $S^1$ manifold. So, at the stellar interior and the exterior region, the topology is $S^2 \times S^1$. 
\item From previous arguments, the equation \eqref{EcuacionReducida} at the stellar surface $\Sigma$, defined by $r=\bar{R}$, must satisfy
\begin{equation}\label{secondform}
    \left[\bar{G}_{\mu\nu}r^{\nu}\right]_{\Sigma}=0, 
\end{equation}
where $r_{\nu}$ is an unit radial vector and  $\left[F\right]_{\Sigma}\equiv F\left(r \rightarrow \bar{R}^{+}\right)- F\left(r \rightarrow \bar{R}^{-}\right)$, for any function $F=F(r)$.
Using the equation (\ref{secondform}) and general Einstein's field equations, one finds
\begin{equation}
    \left[\bar{T}_{\mu\nu}r^{\nu}\right]_{\Sigma}=0,
\end{equation}
which leads to
\begin{equation}\label{secondformpr}
    \left[\bar{p}_r\right]_{\Sigma}=0.
\end{equation}
This is so because the radial pressure is zero for the external vacuum Schwarzchild solution. Then, the last equation yields
\begin{equation}
\bar{p}_r(\bar{R})=0.    \label{p4}
\end{equation}
\end{itemize}

\subsection{A 4-- dimensional geometry example: The Buchdahl space--time}  \label{ToyModel}

As stated before, one of the main aims here is to provide a methodology to explore the consequences of representing a 4--dimensional spherical symmetric compact object into an extra dimension using the BS setup. In this opportunity, to see the effects of the reminiscent compactified extra coordinate $z$ on the 4--dimensional manifold, we have choice the well--known Buchdahl space--time \cite{Buchdahl:1959zz}. So, in this section we revisited in short the principal aspects of the Buchdahl solution and how (\ref{warpfactor1}) is altering the main salient features of the model.

\begin{figure}
\centering
\includegraphics[scale=0.65]{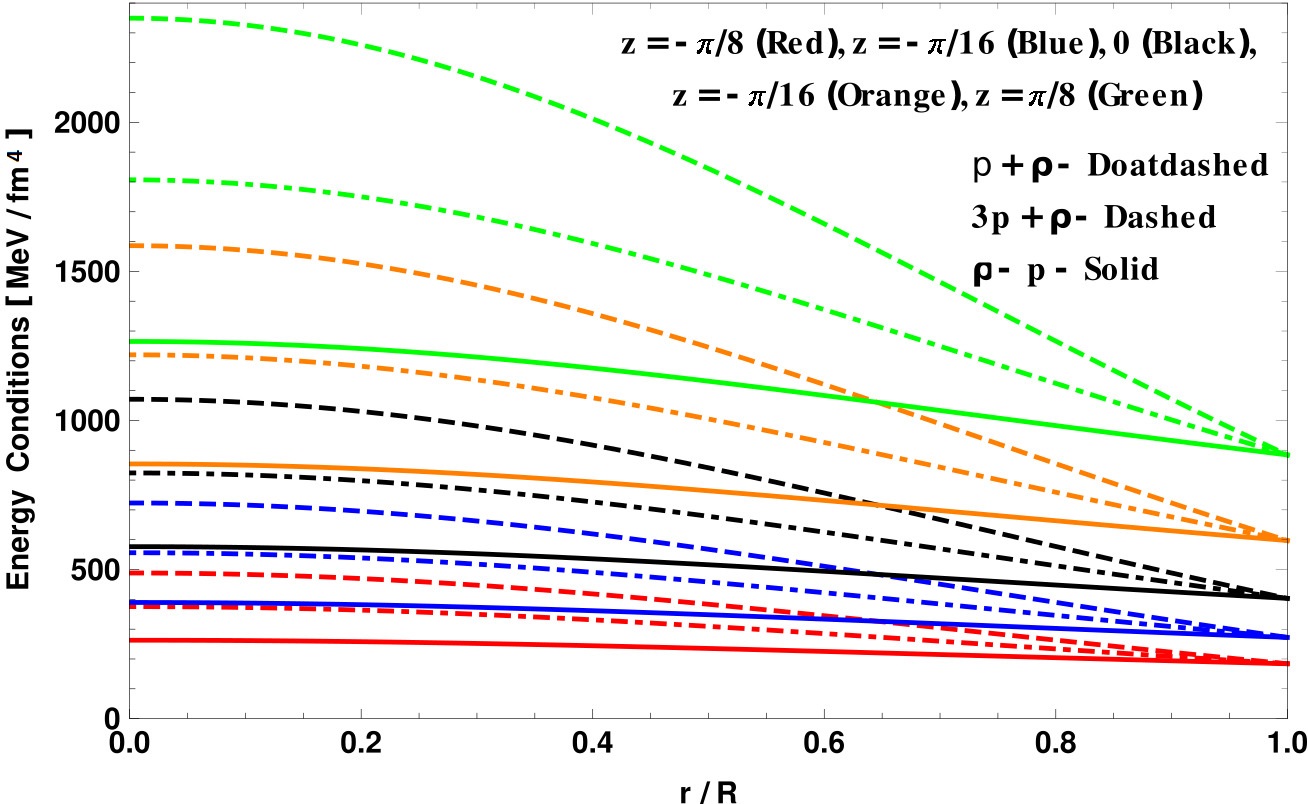}\\
\includegraphics[scale=0.65]{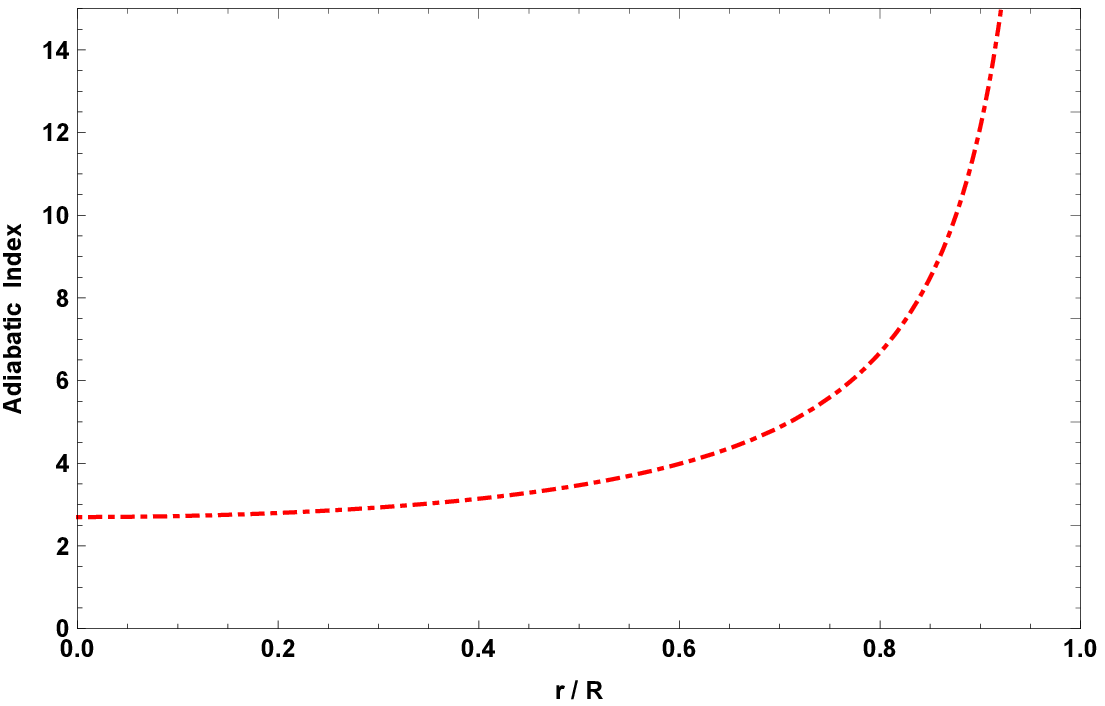}
\caption{\textbf{Upper panel}: The trend of the energy conditions against the dimensionless radial coordinate $r/\bar{R}$. \textbf{Lower panel}: The relativistic adiabatic index versus $r/\bar{R}$, for different values of the extra coordinate $z$.}\label{f2}
\end{figure}

\subsubsection{The Buchdahl metric}
The ansatz corresponds to the well--known Buchdahl space--time, whose metric potentials are given by
\begin{eqnarray}
\hspace{-0.5cm} e^{\nu(r) \over 2} &=& A \left\{B \left(2 c r^2+5\right) \sqrt{2-c r^2}+\left(cr^2+1\right)^{3/2}\right\}\\
\hspace{-0.5cm} \mu(r) &=& \frac{2-c r^2}{2 \left(c r^2+1\right)}, 
\end{eqnarray}
where, $A$ and $B$ are dimensionless integration constants and $c$ is an arbitrary constant with units of $\ell_p^{-2}$. Since this solution is isotropic, we will have $p_r=p_\theta$. Using the field equations \eqref{3tt}--\eqref{3thetatheta}, and previous arguments, we get for the 5--dimensional components of the energy--momentum tensor,
\begin{equation}
\label{density}
8\pi \rho = e^{z/l} K \dfrac{3 \alpha c  \left(c r^2+3\right)}{2 \left(c r^2+1\right)^2},
\end{equation}
\begin{equation}\label{pressure}
\resizebox{1.04\hsize}{!}{$
\begin{split}
8\pi p_r = 8\pi p_\theta = e^{z/l}  K   \frac{9 \alpha c  \left[B \left(2 c^2 r^4-3 c r^2-2\right)+\left(1-c r^2\right) \sqrt{-c^2 r^4+c r^2+2}\,\right]}{2 \sqrt{2-c r^2} \left(c r^2+1\right) \left[B \left(2 c r^2+5\right) \sqrt{2-c r^2}+\left(c r^2+1\right)^{3/2}\right]}~~~
\end{split}$},
\end{equation}
where the constant $\alpha$ has units of $\ell_p^{-2}$. 

It is easy to check that, these 5--dimensional quantities (without bar), which correspond to equation \eqref{EnergiaMomentum1}, are related with the 4--dimensional quantities (with bar) through the relation \eqref{EnergiaMomentum2}, where the function $F \big ( W  (z) \big)=e^{z/l}$ is given by equations \eqref{EcuacionW} and \eqref{warpfactor1}. Moreover, the component $p_z$ of equation \eqref{EnergiaMomentum1}, can be computed from equations \eqref{EnergiaMomentum2} and \eqref{ecuacionp5}. Since the function $F \big ( W  (z) \big)=e^{z/l}$ is positive, and the TEC is negative, hence the component $p_z$ is negative too.

In order to explore the consequences of representing the 4--dimensional Buchdahl space--time into an extra dimension, using the BSs setup, we study ( using a graphical analysis), how the values of the salient thermodynamics properties for different values of $z_0 \in z$ are varying. Following \cite{Aros:2012xi}, a direct computation shows that the induced 4--dimensional energy density and the induced 4--dimensional pressure components for all fixed point $z_0 \in z$, are given by 
\begin{align}
    &\tilde{\rho}=\exp \left (2A(z_0)\right) \bar{\rho}= \rho(r,z_0) \\
    &\tilde{p}_i=\exp \left (2A(z_0)\right) \bar{p}_i =p_i(r,z_0)
\end{align}
with $i=0,1,2,3$.

As can be appreciated, from the above expressions (\ref{density})--(\ref{pressure}), the function (\ref{warpfactor1}) is quite involved in the behavior of the density and pressure. Although, the solution obtained for the function (\ref{warpfactor1}) is simple, the way it affects the thermodynamic properties of the fluid describing the energy--momentum tensor of the 4--dimensional structure, is in principle non--trivial. Of course, different strategies will lead to more complex functions in their shape, perhaps having a more predominant effect on the essential characteristics of the fluid sphere. Next, we shall explore in brief by means of a graphical analysis the impact of the mentioned function (\ref{warpfactor1}) on the configuration.
\begin{figure}
\centering
\includegraphics[scale=0.65]{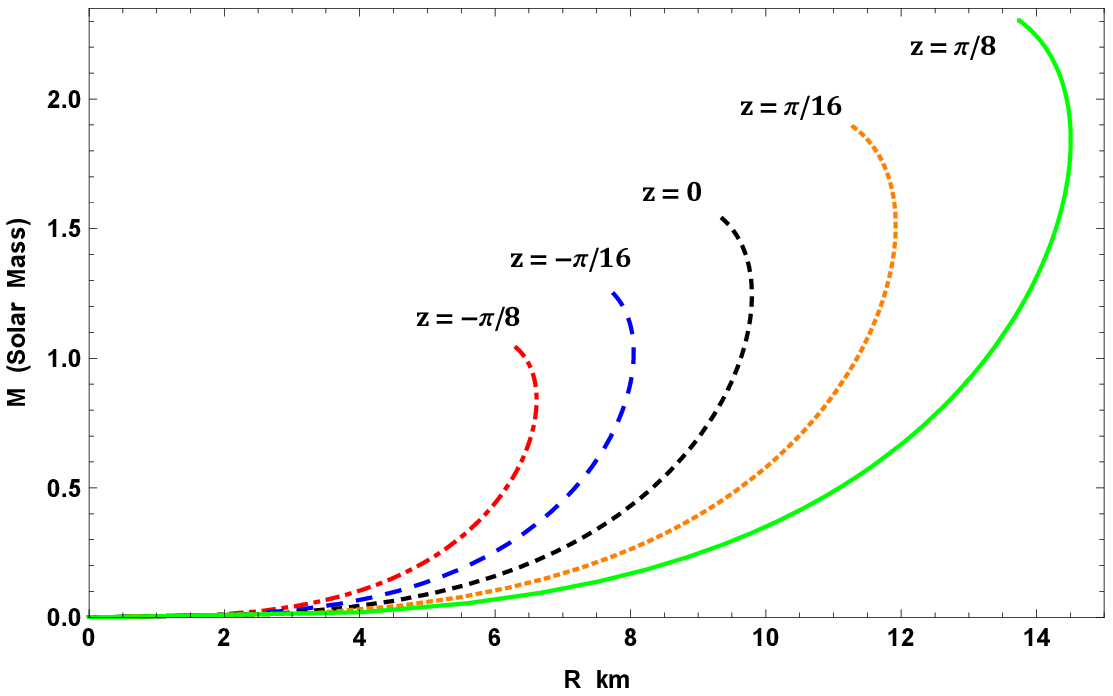}\\
\includegraphics[scale=0.65]{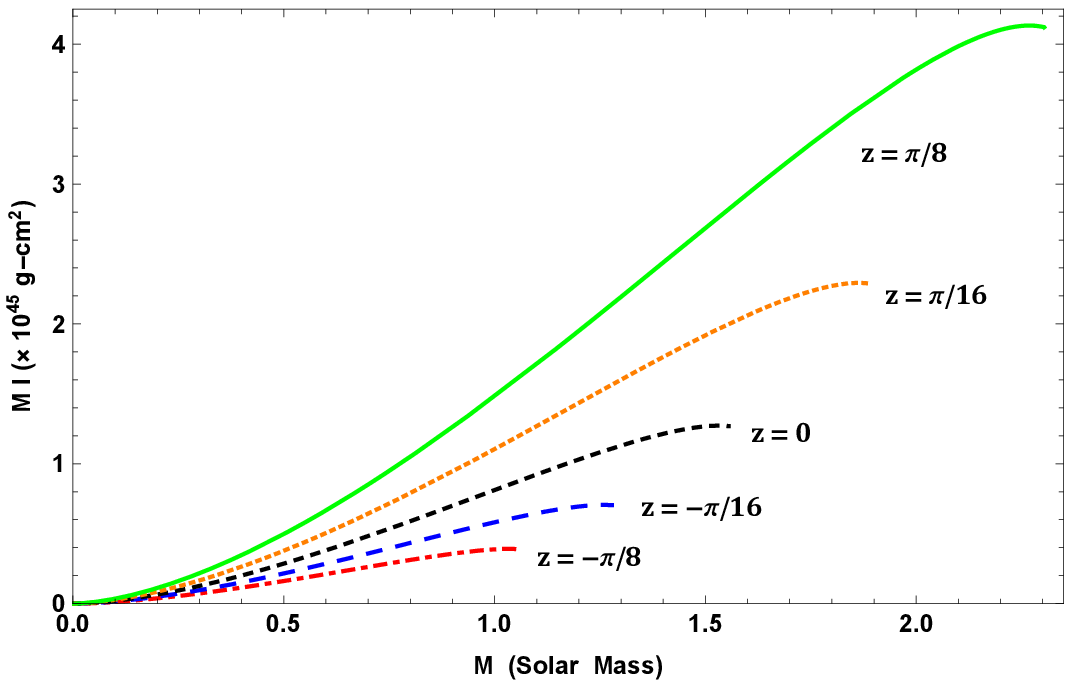}
\caption{\textbf{Upper panel}: The $M$--$\bar{R}$ curve for different values of $z$. \textbf{Lower panel}: The moment of inertia $I$ versus the mass $M$ for different values of the extra coordinate $z$. }\label{f3}
\end{figure}

 \subsubsection{Some properties}
 
As it is well--known, the Buchdhal solution of the Einstein field equations describes a fluid sphere threading by an isotropic matter distribution. Particularly, this space--time meets all the necessary requirements to recreate a realistic toy model for compact stars \cite{Delgaty:1998uy}:
\begin{itemize}
\item The pressure $\bar{p}$ must be a decreasing function such that this one has a finite and positive value at the
centre $r=0$ and it must vanish at the surface $r/\bar{R}=1$.
\item It is required that the energy density be monotonically decreasing towards the boundary $r/\bar{R}=1$.
\item The following energy conditions are required: $\bar{p}+\bar{\rho} \ge 0$, $3\bar{p}+\bar{\rho} \ge 0$, $\bar{\rho}-\bar{p} \ge 0$.
\item The Adiabatic index, $\Gamma$, which determines the stability of the star, must be positive.
\end{itemize}

In order to obtain Figures \ref{f1}--\ref{f3}, previously we have employed junction condition process, Eqs. \eqref{MatchingStar} and \eqref{p4}, to determine the set of constant parameters $\{A,B,c\}$. So, from figures \ref{f1}--\ref{f3} we can highlight the following points

\begin{enumerate}
    \item The Fig. \ref{f1} is displaying the density $\rho$ (upper panel) and pressure \footnote{The scheme presented in Sec. \ref{sec3} concerns a more general situation where the pressures along the radial and tangential directions are unequal \i.e., $p_{r}\neq p_{\theta}$. However, this case corresponds to an isotropic fluid $p_{r}=p_{\theta}$, hence along this section the pressure will be denoted as $p$ only.} $p$ (lower panel) behavior versus the dimensionless radial coordinate $r/\bar{R}$. It is observed that in moving the extra coordinate $z=z_{0}$ from $-\frac{\pi}{8}$ to $\frac{\pi}{8}$, both $\rho$ and $p$ are taking higher values at $r=0$ when $z_{0}\geq0$ and they are decreasing in magnitude when $z_{0}\leq 0$. Then, it is clear that increasing $z_{0}$ in magnitude the core of the structure becomes denser. Furthermore, as the upper panel of Fig. \ref{f2} depicts, the energy conditions are also satisfied, what is more the chance of having an energy--momentum satisfying these conditions increases when $z_{0}$ takes higher values. 
    It is worth mentioning that the units of $\mbox{Energy}/L^4$ can be easily related with Planck units, using $[\mbox{Energy}]=\ell_p^{-1}$ and $[L]=\ell_p$.
    
    \item Interestingly, the relativistic adiabatic index $\Gamma$ is not affected by function $W$ (see lower panel of Fig. \ref{f2}). Indeed, by definition the relativistic adiabatic index $\Gamma$ is given by \cite{10.1093/mnras/239.1.91}
    \begin{equation}
        \Gamma=\frac{\rho+p}{p}\frac{dp}{d\rho}.
    \end{equation}
    So, as the function $W$ is just a global factor, after factorize it from the corresponding expressions for $\rho$ and $p$, it is cancel out
    \begin{equation}
        \Gamma=\frac{\bar{\rho}+\bar{p}}{\bar{p}}\frac{d\bar{p}}{d\bar{\rho}}. \nonumber
    \end{equation}
    Then, in measuring the stability of the model by means of $\Gamma$ this quantity is not altered by the function $W$ (\ref{warpfactor1}). So, in this concern the extra coordinate is not playing any role. 
    \item An important curve in this kind of studies, is the so--called M--$\bar{R}$ curve. As the upper panel of Fig. \ref{f3} exhibits, as $z$ increases in magnitude the mass and radii of the compact object increase too. Moreover, when $z=\frac{\pi}{8}$ the mass of the configuration is above $2M_{\odot}$ whereas for $z=-\frac{\pi}{8}$, the mass is just above $1M_{\odot}$. Then, it is clear that for increasing and positive values of $z$, it is possible to build up more massive and compact objects in comparison with the case where $z=0$. On the other hand, the lower panel in Fig. \ref{f3} is showing the  moment of inertia $I$ versus the mass $M$. This plot has been obtained by considering the following expression
    \begin{equation}
    I=\frac{2}{5}\left[1+\frac{(M/\bar{R})\cdot km}{M_{\odot}}\right]M\bar{R}^{2},    
    \end{equation}
    proposed in \cite{Bejger:2002ty} to covert a static model to a rotating one. As can be seen, the sensitivity of $I$ increases when $z$ moves from $-\frac{\pi}{8}$ to $\frac{\pi}{8}$, reaching its maximum at $z=\frac{\pi}{8}$.
    
It is worth to stress that, when we increase the radius of the compact star the mass content within it also increases and hence we have an increasing curve, however, at a certain radius the mass within it will be large enough, so its gravity will overcome the increase in radius and the additional mass add--on will contracted to a small radius. This kind of nature can seen only in compact stellar systems. Of course, the nature of the $M-\bar{R}$ curve is links with the matter compositions. When the mass increase with radius, the energy density at the center also increases as well as the stiffness corresponding to the equation of state (EoS). However, when the interior density reaches a certain density the stiffness of the EoS suddenly changes due to some kind of phase transitions making the system more compact.

\end{enumerate}
So, as primary conclusion it is evident that the function (\ref{warpfactor1}) is affecting in a positive way the main salient features of the compact structure, specially when the extra coordinate $z$ takes positive values. It is worth mentioning that, to obtain the above results and graphs, the constant appearing in (\ref{warpfactor1}) \i.e., $c$, $\alpha$ and $l$ have been taken to be one.

\section{Concluding Remarks} 

We have provided a way of representing a 4--dimensional stellar distribution using the black string framework. For this, we have chosen a specific form for the 4--dimensional and 5--dimensional line elements and, for the 4--dimensional and 5--dimensional energy--momentum tensor, such that the 4--dimensional and  5--dimensional quantities are related by a function $A(z)$, where $z$ represents  the extra dimension.

One important consequence of the proposed methodology, is that, using the line element \eqref{Elementodelinea1}, the function \eqref{warpfactor1} and the function \eqref{EcuacionW}, the 5--dimensional equations of motion are reduced to the usual 4--dimensional equations of motion \eqref{EcuacionReducida}, which correspond to the line element $d\bar{s}_{4D}^2$. Also, the 5--dimensional conservation equation adopts the form of the usual 4--dimensional conservation equation. 

It is worth mentioning that, although the presented methodology is simple, this reduction form could serve to represent other types of 4--dimensional objects into an extra dimension in future works. This is a not minor point, since in the literature the most $D$--dimensional stellar distributions studied, are those whose topology correspond to $S^{D-2}$ \cite{Chilambwe:2015rra, Dadhich:2016wtb,Molina:2016xeu,Bhar:2014tqa,Paul:2018vho,Estrada:2018vrl}. For example, the topology of a 5--dimensional stellar distribution corresponds to a $S^3$. However, our methodology shows a new way for representing 4--dimensional stellar distributions (whose topology is $S^2$) into an extra compact dimension (whose topology is $S^1$), using the black string setup. So, the topology of the 5--dimensional solution is $S^2 \times S^1$ instead the usual $S^3$ topology.

The schematic representation of our model follows the black string setup, see figures \ref{FigSlides},\ref{FigCilindro} and \ref{FigKKBS}. However, in our case, the horizontal direction is representing the radial coordinate of a 4--dimensional stellar distribution, instead the radial coordinate of a black hole solution as occurs in the usual black strings models.   

In the present model, due to the action of the function $A(z)$, the 5--dimensional geometry no longer has traslational symmetry along the  compact coordinate $z$, as occurs for the Kaluza--Klein black string. On the other hand, under our assumptions the sign of the pressure along the extra dimension is given by the trace of the 4--dimensional energy--momentum tensor. 

In the subsection \eqref{Matching} we have provide the matching condition for our form of representation of $4D$ stellar distribution using the black string style, which are reduced to the usual four dimensional matching condition.

In order to explore the consequences of representing 4--dimensional stellar distribution into an extra dimension in the black string framework, we have characterized the 4--dimensional Buchdahl geometry \cite{Buchdahl:1959zz} as toy model. One of the most relevant characteristics of this scheme, is that allows the construction of more stable and compact objects. This is quite relevant from the astrophysical point of view, because it is clear that this mechanism works as a mass generator. Particularly, in this case, as the coordinate $z=z_{0}$ moves from $-\frac{\pi}{8}$ to $\frac{\pi}{8}$ the mass of the object increases, leading to more compact configurations than in the normal ($z=0$) scenario. We wish to emphasize that this protocol is not a tool to find analytical solutions to 4--dimensional Einstein's field equations, but rather to establish guidelines for determining how a 4--dimensional solution is immersed in an extra dimension, when the approach at the black string style is applied.

For a fixed value of $z=z_0$, inside the stellar distribution, the topology corresponds to the product between the Buchdahl geometry and $S^1$. Outside the stellar distribution the topology corresponds to the product between the Schwarzschild geometry and $S^1$. So, inside and outside of the stellar distribution, the topology is $S^2 \times S^1$.

\bibliography{mybib}

\end{document}